\documentclass[epsf]{aa}
\usepackage{natbib}
\usepackage{epsfig}

\def\Msol{M$_\odot$}             %Msun
\def\Rsol{R$_\odot$}             %Msun

\begin{document}
%%%%%%%%%%%%%%%%%%%%%%%%%%%%%%%%%%%%%%%%%%%%%%%%%%%%
\title{The transiting planet OGLE-TR-132$b$ revisited with new spectroscopy and deconvolution photometry\thanks{Based on data collected with the FORS2 imager at the VLT-UT4 telescope (Paranal Observatory, ESO, Chile)  in the programme 273.C-5017A, with the SUSI2 imager at the NTT telescope (La Silla Observatory, ESO, Chile) in the programme 075.C-0462A, and with the UVES spectrograph at the VLT-UT2 telescope (Paranal Observatory, ESO, Chile) in the programme 076.C-0131.}}

\author{ Pont F.$^1$, Gillon M.$^{1, 2}$, Moutou C.$^3$, Santos N. C. $^{1, 4, 5}$, Bouchy F.$^{6, 7}$,  Mayor M.$^1$,  Melo C. $^8$, D. Queloz$^1$, S. Udry$^1$}

\offprints{frederic.pont@obs.unige.ch}
\institute{$^1$  Observatoire de Gen\`eve, 51 Chemin des Maillettes,1290 Sauverny,
    Switzerland\\
$^2$ Institut d'Astrophysique et de G\'eophysique,  Universit\'e
  de Li\`ege,  All\'ee du 6 Ao\^ut 17,  4000 Li\`ege, Belgium \\
$^3$ LAM, Traverse du Siphon, BP8, Les Trois Lucs, 13376 Marseille
  cedex 12, France\\
$^4$ Centro de Astronomia e Astrof{\'\i}sica da Universidade de Lisboa,
  Observat\'orio Astron\'omico de Lisboa, Tapada da Ajuda, 1349-018
  Lisboa, Portugal\\
$^5$   Centro de Geofisica de \'Evora, Rua Rom\~ao Ramalho 59,
  7002-554 \'Evora, Portugal\\
$^6$ Observatoire de Haute-Provence, 04870 St-Michel l'Observatoire, France\\
$^7$ Institut d'Astrophysique de Paris, 98bis Bd Arago, 75014 Paris, France\\
$^8$ European Southern Observatory, Casilla 19001, Santiago 19, Chile\\
%$^8$ Affiliation 8
}
\date{Received date / accepted date}
   \authorrunning{F. Pont et al.}
   \titlerunning{The radius of the transiting planet OGLE-TR-132b}
%%%%%%%%%%%%%%%%%%%%%%%%%%%%%%%%%%%%%%%%%%%%%%%%%%%%
\abstract{OGLE-TR-132b is transiting a very metal-rich F dwarf about 2000 pc from the Sun, in the Galactic disc towards Carina. It orbits very close to its host star and has an equilibrium temperature of nearly 2000 K. Using rapid-cadence transit photometry from the FORS2 camera on the VLT and SUSI2 on the NTT, and high-resolution spectroscopy with UVES on the VLT, we refine the shape of the transit lightcurve and the parameters of the system. In particular, we improve the planetary radius estimate, $R=1.18 \pm 0.07 R_J$ and provide very precise ephemerids, $T_{\mathrm tr}=2453142.59123 \pm 0.0003$ and $P=1.689868 \pm 0.000003$. Our results give a slightly smaller and lighter star, and bigger planet, than previous values. We outline a normalisation problem with some implementations of differential image photometry, that is solved by the use of deconvolution photometry.
}
\maketitle
%%%%%%%%%%%%%%%%%%%%%%%%%%%%%%%%%%%%%%%%%%%%%%%%%%%%

\section{Introduction}

Ten transiting extrasolar planets are now known, and they have proved invaluable in the study of hot Jupiter structure and evolution \cite{Charbonneau}. OGLE-TR-132b is a very short-period hot Jupiter (P=1.69 days) orbiting a very metal-rich F dwarf. The photometric transit signal was detected by Udalski et al. \cite{Udalski} from the OGLE planetary transit search on the Warsaw 1.3m telescope at Las Campanas, Chile. The spectroscopic orbit was measured by Bouchy et al.  \cite{Bouchy}, establishing the planetary mass of the transiting companion. High-accuracy photometry of the transit was reported by  Moutou et al. (2004, hereafter paper I) using the FORS2 camera on the VLT.

OGLE-TR-132b occupies the edge of parameters' space in several respects among the ten known transiting planets: it has a very short period and a rather large primary, within the uncertainties it is therefore probably the known gas giant receiving the most flux from its host star. This later has very high metallicity. The combination of these two factors, for instance, make OGLE-TR-132b a crucial object for the relation between star metallicity and planet core size proposed by Guillot et al. \cite{Guillot}.  Therefore, refining the knowledge of its parameters is useful to provide constraints for hot Jupiter structure, formation and evaporation scenarios.

Given the faintness of OGLE-TR-132 (I=15.7), large telescopes are necessary for the photometric and spectroscopic measurements.

In this paper, we present new high-resolution, high-S/N spectroscopic observations used to refine the parameters of the host star of OGLE-TR-132b, and high-accuracy photometric coverage of two transits, one year apart, to improve the accuracy of the planetary radius and the orbital ephemerids. The data were obtained with UVES on the VLT for the spectroscopy, and with FORS2 on the VLT and SUSI2 on the NTT for the photometry.

The methods to analyze the transit shape, estimate the uncertainties in the presence of photometric systematics and derive the stellar and planetary parameters are the same as in our previous studies of OGLE transiting planets (Bouchy et al. 2004; 2005; Pont et al. 2004; 2005; Moutou et al. 2004; Santos et al. 2006; Gillon et al. 2006) and will only be repeated briefly here when necessary. The reader is referred to those papers for details.

\section{Observation and reduction}

\subsection{Photometry}
\label{Sphot}

\subsubsection{Observations}

The VLT/FORS2 data used here were presented in Paper I. NTT observations were obtained on April 19th, 2005 on the SUSI2 camera (programme 075.C-0462A). 272 exposures were acquired in a 5.4$\arcmin$ $\times$ 5.4$\arcmin$ field of view. The exposure time was 60 s, while the read-out time was 22 s. The measured seeing varies between 0.75$\arcsec$ and 1.38$\arcsec$. The $R\#813$ filter was used for all observations. We used SUSI2 with a 2 $\times$ 2 pixel binning to get a good spatial and temporal sampling at the same time. The binned pixel size is 0.16\arcsec. The air mass of the field decreases from 1.31 to 1.19 then grows to 1.54 during the sequence. The quality of the night was not photometric (large transparency variations).

The frames were debiassed and flatfielded with the standard ESO pipeline. 

\subsubsection{Reduction}

We used the deconvolution-based  photometric reduction method DECPHOT\footnote{DEConvolution-based PHOtometry. We introduce this acronym here.}, described in Gillon et al. \cite{Gillon} and Magain et al. (2006) to reduce both VLT and NTT data. This method relies on the partial deconvolution of a set of images to the same higher resolution, leading to the minimisation of the systematic effects due to seeing variations along the run, the accurate determination of the sky background, and the detection of faint blended sources undetected in the original images. Furthermore, the Point-Spread Function ($PSF$) is determined \emph{simultaneously} to the deconvolution of the images using all the sources present in the field. This $PSF$ determination does not rely on the presence of any isolated star in the field, it is thus well suited for crowded fields photometry. With DECPHOT, the analysis of a set of images is divided into two parts. First, a master high-S/N image (a combination of the best seeing images) is analyzed to detect all the sources and their position in the field. In a second step, the astrometry is kept fixed and the entire set of images is analyzed, allowing to obtain an accurate photometry for every sources, even faint blended ones. 

In Paper I, the VLT data were reduced using the Difference Image Analysis (DIA) ISIS software (Alard et Lupton 1998; Alard 1999; 2000). This algorithm relies on the use of a high-S/N reference image which is convolved to the seeing of the analyzed image and subtracted, leading to a difference image. The flux in this image is then measured by aperture or profile photometry. This kind of reduction method is generally presented as very suitable for crowded fields photometry, as variability is measured on a nearly empty surface, while other reduction methods aim to measure the total flux of objects which could be strongly blended \cite{Udalski2}. It allowed to obtain in many cases residuals deviations very close the photon noise limit (see e.g. Paper I; Holman et al. 2005).  In the case of our first analysis of the VLT data (Paper I), the flux in the reference image and in the difference images was measured with aperture photometry, using the same aperture for every images. It appeared from our re-analysis of the data that with this reduction method, a tradeoff has to be chosen between the relative accuracy of the lightcurve, and the reliability of the absolute amplitude of the flux variation, as illustrated on Fig. 1. This figure shows clearly that using smaller apertures leads to an artificial reduction of the amplitude of any variation in the lightcurve, and thus also of the standard deviation, an effect already reported by Hartman et al. (2004). This point could contribute to explain the fact that OGLE-TR-10 light curves obtained by Holman et al. (2005) using a similar reduction method have a small standard deviation and a depth significantly smaller than the one present in OGLE-III light curves \cite{Konacki}. 
 
\begin{figure}
\label{fig:a}
\centering                     
\includegraphics[width=9.0cm]{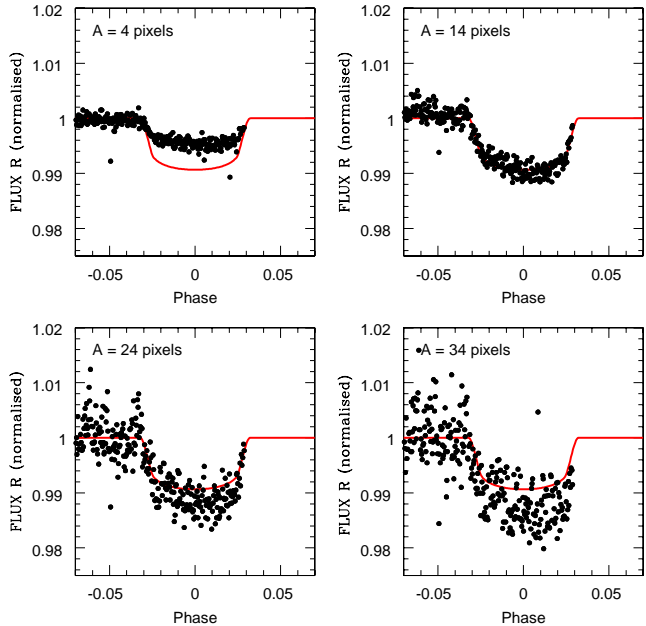}
\caption{VLT/FORS2 lightcurves for OGLE-TR-132b, obtained with ISIS and aperture photometry, using 4 different apertures $A$. The best-fit transit curve obtained after reduction with DECPHOT is superimposed in red. The deviations before the transit are 1.0 mmag ($A$ = 4 pixels), 1.9 mmag ($A$ = 14 pixels), 4.4 mmag ($A$ = 24 pixels) and 5.6 mmag ($A$ = 34 pixels).}
\end{figure}

Wiith DIA methods, one uses the stellar flux measured in the reference image, i.e. the same for every images of the run, to convert the differential flux units into magnitudes. Any difference between the fraction of the stellar flux measured in difference images and in the reference image is thus able to artificially modify the lightcurve. With the reduction method used in Paper I, it was definitely the case: the fraction of the stellar flux within the aperture was larger in the average difference image than in the reference image, as this later is the best-seeing image of the run. We also notice that any inaccuracy of the analytical kernel connecting the reference image to the other ones is able to lead to a similar systematic error. 

When trying to detect variables and transits, a great accuracy on the amplitude is not crucial, and the DIA reduction can be optimized to give the lowest possible noise on the photometry. It must also be pointed out that DECPHOT takes several orders of magnitude more computer time to operate than DIA. However, when measuring the parameters of a planetary transit, it is desirable to attain the lowest noise possible while still retaining a correct normalisation of the signal amplitude. In this case,  the deconvolution method, which does not require a trade-off between the two factors, can offer an improvement. 

\subsection{Spectroscopy}

High-resolution spectra of OGLE-TR-132 were obtained using the UVES spectrograph at the VLT-UT2 Kueyen telescope (program ID\,076.C-0131). Eight exposures of 3000 seconds each were done between December 2005 and January 2006. Each individual spectrum was then combined using the IRAF\footnote{IRAF is distributed by National Optical Astronomy Observatories, operated by the Association of Universities for Research in Astronomy, Inc.,under contract with the National Science Foundation, U.S.A.} {\tt scombine} routine. The total S/N obtained is close to 100, as measured directly from small spectral windows with no clear spectral lines in the region near 6500\AA.

As in our previous spectroscopic study of other OGLE stars (Santos et al. 2006), the CCD was read in 2x2 bins for each exposure to reduce the readout noise and increase the number of counts in each bin. Similarly we opted for using a slit width of 0.9 arcsec, which provides a spectral resolution R=$\lambda$/$\Delta\lambda$$\sim$50\,000. The observations were made using the Dichroic 390+580 mode. The red portion of the spectra (used in this paper) covers the wavelength domain between 4780 and 6805\AA, with a gap between 5730 and 5835\AA.

As before, particular attention was paid to the orientation of the slit due to the relatively crowded field. The angle was chosen using the images available at the OGLE website\footnote{http://www.astrouw.edu.pl/$\sim$ftp/ogle/index.html}, so that no other star was present in the UVES slit during the observation.

The final spectrum of OGLE-TR-132 was used to derive stellar parameters and iron abundances. These were derived in LTE using the 2002 version of the code MOOG (Sneden 1973)\footnote{http://verdi.as.utexas.edu/moog.html} and a grid of Kurucz Atlas plane-parallel model atmospheres (Kurucz 1993). The whole procedure is described in detail in Santos et al. (2006, and references therein) and is based on the analysis of 39 \ion{Fe}{i} and 12 \ion{Fe}{ii} weak lines and on imposing excitation and ionization equilibrium. The resulting parameters are listed in Table\,1. 

\section{Results}

Figure 2 presents the lightcurves from the VLT and  NTT data reduced with DECPHOT. The deviation of the VLT light curve before the transit is 1.12 mmag, while the mean photon noise is 1.04 mmag. The deviation of the NTT lightcurve after the transit is 1.60 mmag, while the mean photon noise is 1.30 mmag. The smaller accuracy of the NTT lightcurve is due to the atmospheric conditions during the observations. 

\begin{figure}
\label{fig:c}
\centering                     
\includegraphics[width=9.0cm]{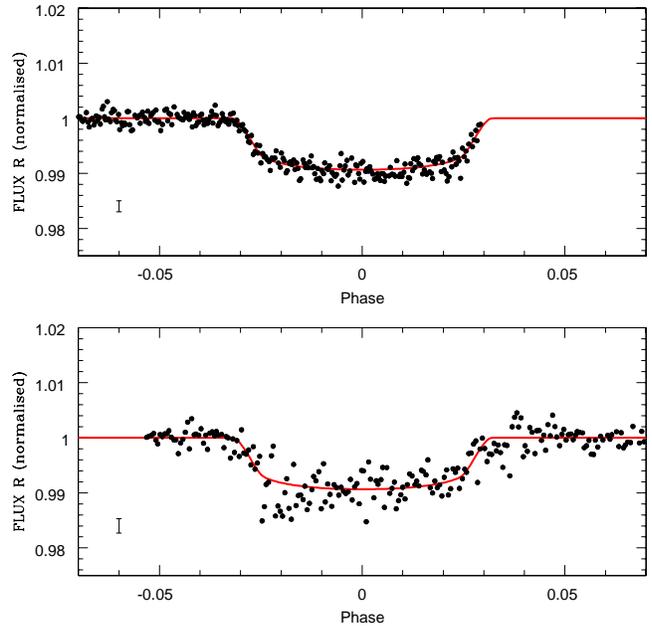}
\caption{VLT/FORS2 (\emph{top}) and NTT/SUSI2 (\emph{bottom})  lightcurves  folded at the best phase and period. The best-fit transit curve is superimposed in red. }
\end{figure}

Figure 3 shows that  no close companion contaminates the $PSF$'s core of  OGLE-TR-132 in the deconvolved best-seeing VLT image. This relative isolation allowed us to reduce the VLT data with aperture photometry and to check the agreement between the resulting light curve and the one obtained with DECPHOT. For this purpose, we used the DAOPHOT aperture photometry software (Stetson, 1987) with a rather small aperture (3 pixels) to avoid any dilution of the transit signal and noise increase due to the East faint blending companion. Figure 4 compares the best light curve obtained with the one from DECPHOT. The agreement between both light curves is excellent. This demonstrates that, as our deconvolution-based method does not require any conversion of the measured flux to obtain the final photometry (the master image is only used with DECPHOT to obtain an accurate astrometry, see Section~\ref{Sphot}), there is no risk for the resulting light curves to be artificially modified. The accuracy obtained with DECPHOT is significantly better than the one obtained with aperture photometry, which is comparable to the one obtained with ISIS/aperture photometry (paper I), if we correct this later by the appropriate scale factor (standard deviation $\sim$ 1.5 mmag).

\begin{figure*}
\label{fig:bbc}
\centering                     
\includegraphics[width=12.0cm]{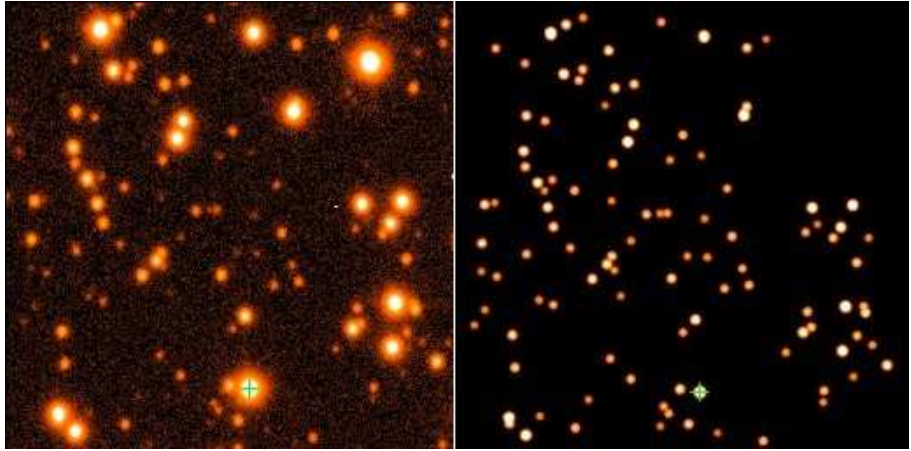}
\caption{\emph{Left}: OGLE-TR-132 (marked with a cross) in a 256 pixels $\times$ 256 pixels sub-image (0.51 $\arcmin$ $\times$ 0.51 $\arcmin$) from the best seeing VLT/FORS2 image of the run ($top$ = North, $left$ = East). \emph{Right}: deconvolved image. }
\end{figure*}

\begin{figure}
\label{fig:uuu}
\centering                     
\includegraphics[width=9.0cm]{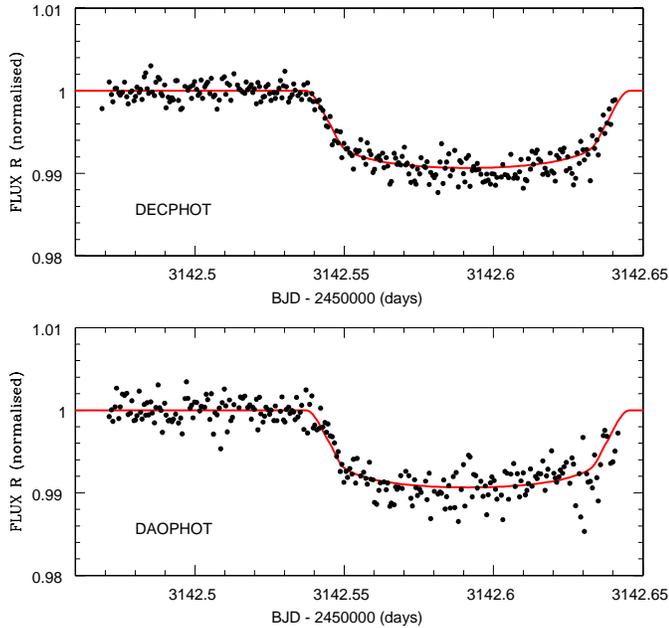}
\caption{VLT/FORS2 lightcurve for OGLE-TR-132b, obtained with DECPHOT (\emph{top}) and DAOPHOT (\emph{bottom}). The best-fit transit curve is superimposed in red.}
\end{figure}

We notice also the excellent agreement between the depths of the two different nights, reduced independently. Such a stability in the depth normalisation was also found for two different transits in the similar case of OGLE-TR-113 (Gillon et al. 2006). In the context of measuring parameters for transiting planets, this is the primary advantage of DECPHOT.

Table~\ref{spectro} lists the spectroscopic parameters obtained from our analysis of the UVES spectra.  These values are compatible with previous estimates by Bouchy et al. (2004) from lower S/N spectra, while being more precise. In particular, the very high metallicity of the host star of OGLE-TR-132b is confirmed. The temperature is near the lower edge of the error bar of the previous estimate, and therefore favor a slightly smaller primary.

\section{Parameters of the OGLE-TR-132 system}

\begin{table}
\begin{tabular}{l l } \hline \hline
{\it Star}  & \\
 & \\
 $T_{\mathrm eff}$  [K]& 6210 $\pm$59\\
 $\log g$ [cms]& 4.51 $\pm$ 0.27 \\
 $\left[ F\! e/H \right] $ & 0.37 $\pm$ 0.07 \\
 & \\
 Mass [\Msol ] & 1.26 $\pm$  (0.03) \\
 Radius [\Rsol ]& 1.34 $\pm$  0.08 \\
 & \\
 {\it Transit} & \\
 & \\
 Impact parameter & 0.53  $\pm$  0.09 \\
 Radius ratio & 0.0916 $\pm$   0.0014 \\
 $V_T/R$ & 17.79 $\pm$ 1.15 \\
 & \\
 {\it Planet} & \\
 & \\
Mass [$M_J$]& 1.14 $\pm$  0.12 \\
Radius [$R_J$]& 1.18 $\pm$  0.07\\
Density [g cm$^{-3}$]& 0.86$^{+0.28}_{-0.21}$ \\
Period [days]& 1.689868 $\pm$ 0.000003 \\
$T_{\mathrm tr}$ [BJD] & 2453142.5912 $\pm$ 0.0003  \\ \hline
\end{tabular}
\caption{Parameters for the OGLE-TR-132 system, host star and transiting planet.The uncertainty on the stellar mass is given in bracket to emphasis the fact that it does not include the uncertainties on stellar evolution models, that could make it higher.} 
\label{spectro}
\label{transit}
\label{param}
\end{table}

\subsection{Transit parameters}

We have fitted transit lightcurves, calculated from Mandel \& Agol (2002), on the VLT and NTT lightcurves independently, then on the joint lightcurves, with quadratic limb-darkening coefficients from Claret (2000). The results are given in Table~\ref{transit}.  Like in Bouchy et al. (2005), we use the radius ratio, the impact parameter (minimum distance from the center of the planet to the center of the star divided by the star's radius), and $V_T/R$ (the transverse velocity of the planet at the center of the transit in terms of the star's radius). These three parameters are independent of any assumption on the size and mass of the primary.  We have used the circular residual shift method explained in Moutou et al. (2004) to estimate error bars on the parameters that take into account possible trends in the photometry.  We note that the differences between the values fitted independently on each transit are compatible within the uncertainty estimates of the combined fit, which may indicate that the uncertainties are realistic.

If the sum of the two limb-darkening coefficients is left as a free parameter, it converges to a value that is 0.1 lower than the coefficients in Claret (2000), with an uncertainty of similar amplitude. In our adopted solution, we fix the limb-darkening coefficients to their tabulated value.

\subsection{Host star parameters}

To estimate the mass and radius of the host star from its spectroscopic parameters, some additional assumption is needed. As in previous papers (see Introduction), we assume that the star follows the stellar evolution models of Girardi et al. (2002) and we combine the constraints from the spectroscopic and transit parameters by Maximum Likelihood. We point out that the derived mass and radius of the star, and therefore that of the planet, will be correct only to the extent that these models are accurate. For such a high metallicity, significant discrepancies are not excluded. However,  our high-accuracy photometric data helps lifting the degeneracy between the star's radius and the impact parameter, providing a constraint on the star's radius that is independent of the models.

The 1-sigma interval for the stellar evolution models from the Maximum Likelihood solution is 0-2 Gyr. The upper limit is slightly higher than that estimated by Moutou et al. (2004), 1.5 Gyr, because of the lower temperature we obtain for the primary. From a measurement of the Calcium emission line for this star from the same UVES spectra, Melo et al. (2006) estimate that is must be older than 0.5 Gyr. Therefore, we estimate that OGLE-TR-132 has an age in the 0.5 - 2 Gyr range.

\subsection{Planetary mass and radius}

The planetary radius is simply obtained from the stellar radius and the radius ratio for the photometry. The planetary mass is obtained from the stellar mass and the radial velocity semi-amplitude. Although we had no new radial velocity data compared to Bouchy et al. (2004), our data also lead to a slight change in the estimated mass because of the improved ephemerids. 

Table~\ref{param} gives the derived values for the stellar and planetary mass and radius, and the planet density.

\subsection{Ephemerids}

The uncertainty on the central epoch is about 25 seconds for the VLT transit and 50 seconds for the NTT transit. If the transits are strictly periodic, this translates into an uncertainty of $3 \cdot 10^{-6}$ days (0.26 seconds) on the period.

By fitting the transit shape obtained from Table~\ref{transit} on the original OGLE data, with only the epoch and period as free parameters, we obtain $P=1.68985 \pm 0.00001$ from the OGLE photometry (uncertainty assuming white noise only). This is in agreement with the difference between the epoch of the VLT and NTT transit, and shows that any time delay effect between the two does not exceed an amplitude of $\sim$ 6 minutes. Using eq. 33 from Agol et al. (2005) and assuming that the phase difference between  both transits leads to a maximal time delay, this maximal delay corresponds to a $4 M_{\oplus}$ exterior planet in 2:1 mean-motion resonance. A lighter planet in this configuration cannot be excluded, neither a heavier one in another orbital and/or phase configuration. 

Figure 5 shows the OGLE data folded at the best phase and period from the VLT and NTT lightcurves.

\begin{figure}
\label{fig:www}
\centering                     
\includegraphics[width=8.0cm]{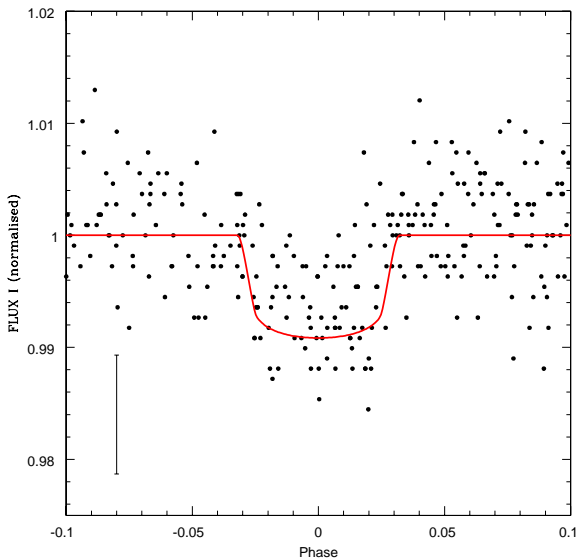}
\caption{OGLE data folded at the best phase and period from the VLT and NTT lightcurves, with the best-fit transit curve superimposed in red.}
\end{figure}

\subsection{Discussion}

The present work provides a very reliable description of the  shape of the planetary transit lightcurve of OGLE-TR-132b, and highly accurate ephemerids that will allow to predict accurate transit timings for the many years. The spectroscopic parameters of the host star were also significantly improved.
Compared to previous estimates, the new data indicate a slightly cooler host star, confirmed as very metal-rich. The transit depth is found deeper than indicated by Moutou et al. (2004). As detailed in Section~\ref{Sphot}, this was due to a normalisation problem with DIA photometry, solved by our use of deconvolution and measurement of two different transits with different instruments.

The effect of a cooler star, leading to a smaller estimated primary radius, and of a deeper transit, leading to a larger radius ratio, compensate to first order, so that our value of the planetary radius is only slightly larger than that reported in Moutou et al. (2004), and compatible with the error bars. The cooler temperature also implies a decrease in the estimated stellar mass, with a corresponding decrease in the planetary mass, also within the error bars (the uncertainty on the planetary mass is dominated by that of the amplitude of the radial velocity orbit).

The baseline provided by our lightcurves for the detection of transit timing delay effects, due to the presence of a second planet or a satellite, is of the order of a minute. This is less constraining than in the similar case of OGLE-TR-113 (Gillon et al. 2006) because the photometric accuracy is similar and the transit is much shallower. In this case, a several Earth-mass satellite would not be detectable even in a favourable configuration (obviously, any satellite is also very unlikely given the very small orbital distance).

\begin{acknowledgements} 
The authors thank the ESO staff on the VLT and NTT telescopes for their diligent and competent execution of the observations. M.G. acknowledges the PLANET network for its support in the form of a fellowship (reference HPRM-CR-2002-00308). Support from the Funda\c{c}\~ao para a Ci\^encia e a Tecnologia (Portugal) to N.C.S. in the form of a fellowship (reference SFRH/BPD/8116/2002) and a grant (reference POCI/CTE-AST/56453/2004) is gratefully acknowledged. 
\end{acknowledgements}

\end{document}